\documentclass[nohyper,11pt,letterpaper]{JHEP3}
\usepackage[dvips]{epsfig}
\usepackage{amsmath}
\usepackage[active]{srcltx}

\newcommand{\be}{\begin{equation}}
\newcommand{\ee}{\end{equation}}
\newcommand{\bea}{\begin{eqnarray}}
\newcommand{\eea}{\end{eqnarray}}
\newcommand{\ba}{\begin{eqnarray}}
\newcommand{\ea}{\end{eqnarray}}

\newcommand{\beq}{\begin{equation}}
\newcommand{\eeq}{\end{equation}}
\newcommand{\beqa}{\begin{eqnarray}}
\newcommand{\eeqa}{\end{eqnarray}}
\newcommand{\beqar}{\begin{eqnarray*}}
\newcommand{\eeqar}{\end{eqnarray*}}

\newcommand{\reef}[1]{(\ref{#1})}

\def\nc {N_\mt{c}}

\def\t6 {T_\mt{D6}}

\newcommand{\mt}[1]{\textrm{\tiny #1}}

\newcommand{\nq}{n_q}

\newcommand{\overlrarrow}[1]{\vbox{\ialign{##\cr\cr
                  \leftrightarrowfill\crcr\noalign{\kern-1pt\nointerlineskip}
                  $\hfil\displaystyle{#1}\hfil$\crcr}}}
\newcommand{\un}{u_{0}}

\newcommand{\tilf}{\tilde{f}}

\title{On holographic phase transitions at finite chemical potential}

\author{Shunji Matsuura \\
Perimeter Institute for Theoretical Physics,
Waterloo, Ontario N2L 2Y5, Canada \\
Department of Physics, University of Tokyo,
7-3-1 Hongo, Bunkyoku, Tokyo\\
\ \ 113-0033, Japan\\
\\E-mail: \email{smatsuura@perimeterinstitute.ca}}

\date{\today}

\abstract{
Recent Holographic studies have shown that N=4 super Yang-Mills theory coupled to fundamental matter with finite chemical potential undergoes a first order phase transition.
In this paper, we study $N_f$ D6 probe branes with or without electric field on it in the black D4 brane background compactified
on a circle with supersymmetry breaking boundary condition.
At energy scales much lower than the compactification scale, 
the dual gauge theory is effectively four dimensional non-supersymmetric
SU($N_c$) Yang-Mills theory coupled to fundamental matter with or without baryon number charge. 
Within the supergravity approximation, the decoupling of the Kaluza-Klein modes is not fully realized.
For chemical potential $\mu_b<N_cM_q$ there is a line of first order phase transitions 
from stable meson phase to unstable meson phase.
On the other hand for $\mu_b>N_cM_q$ there is no phase transition
and mesons are unstable.
A peculiar and interesting property of this system is that for a 
certain range of chemical potential $\mu_b<N_cM_q$,
a new phase transition appears in the unstable meson phase.
This phase transition is characterized by a discontinuous 
change of unstable meson lifetime.
}

\begin{document}



\section{Introduction}

Thermal phase structure of
strongly coupled SU($N_c$) super Yang-Mills theory at finite chemical potential and finite temperature may be studied by using holographic duality \cite{juan, bigRev}.
The holographic method provides a powerful framework to study a broad class of large $N_c$, strongly coupled gauge theories with a small number $N_f$ of fields in the fundamental representations
\cite{johanna,Kruczenski:2003uq,Mateos:2006nu,Mateos:2007vn,visco,recent}.
The gravity dual of these gauge fields appear as $N_f$ probe Dq-branes on
the near horizon geometries of $N_c$ black Dp-branes.

Particularly interesting physics appears when the system undergoes a confinement/de- confinement 
phase transition.
Above this phase transition temperature $T_{dec}$, the gluons and the adjoint matters
are deconfined and the dual geometry contains a black hole \cite{Witten:1998zw}.
Since our main understanding of QCD largely comes from the behaviour of fundamental matter,
it is interesting to ask how they behave around the deconfinement phase transition.
The thermal properties of fundamental matter in the deconfinement phase can be 
extracted by studying probe D-branes in this black hole background.

In \cite{Mateos:2006nu, Mateos:2007vn}, it was show that at zero baryon density, 
fundamental matter undergoes a first order phase transition at $T_{fun}$.
In the lower temperature phase, the probe D-brane sits entirely outside the black hole 
horizon, which is called a Minkowski embedding (see Figure \ref{config}), and
the mass spectrum of meson in this phase is discrete and has
a mass gap\cite{Mateos:2006nu,Mateos:2007vn}.
In the higher temperature phase, a part of the probe D-brane fall through the black hole horizon, which is called a black hole embedding, and  
the mass spectrum of meson in this phase is continuous and gapless, characterized by quasinormal modes
\cite{Mateos:2007vn,spectre,hoyos}.
We emphasize that this is in the deconfinement phase.
That means that after the gluons and the adjoint matters are deconfined, the fundamental matter is still in bound states at temperature $T_{dec}<T<T_{fun}$.
The meson bound states in the deconfinement phase 
are also illustrated in related models \cite{recent8}.
On the other hand if $T_{fun}$ is smaller than $T_{dec}$,
the confinement/deconfinement phase transition and the meson melting
take place simultaneously.
This phenomenon, the stable meson bound states in a deconfinement phase $T_{dec}<T<T_{fun}$, is also found in lattice QCD\cite{lattice}, which
suggests that bound states of heavy quarks
survive after the deconfinement phase transition up to a few $T_{fun}$.

\FIGURE{
  \includegraphics[width=0.5 \textwidth]{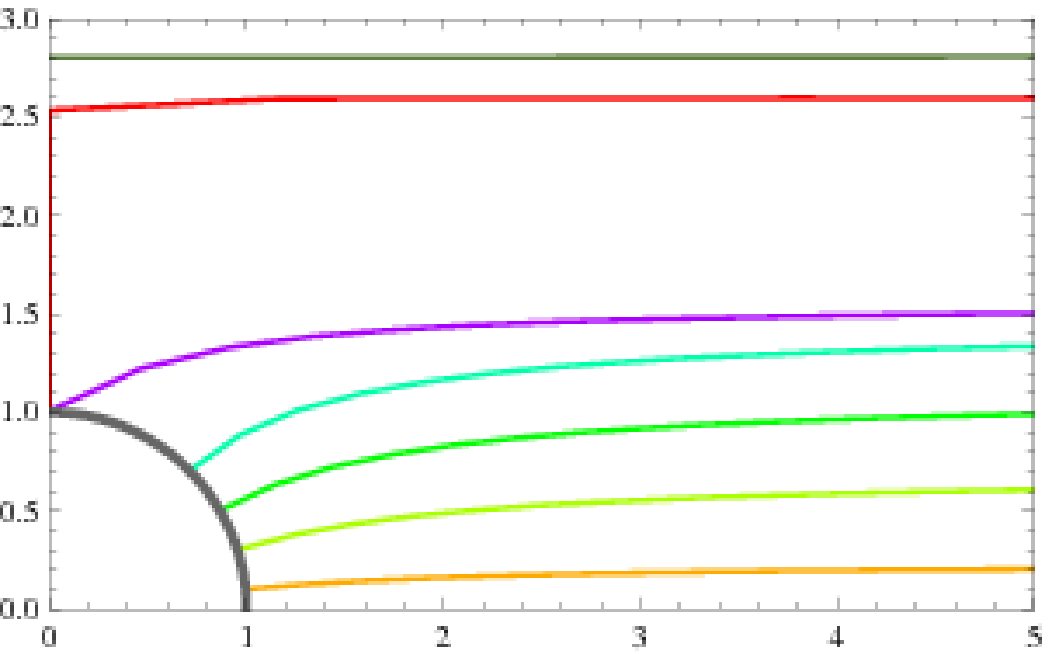}
\caption{Probe D-brane configurations in the presence of the black hole.
Gray circle represents the black hole horizon.
For the black hole embeddings, temperature decreases from the bottom to top.
The dark green colour represents Minkowski embedding which is entirely outside the
horizon.
At lower temperature, the probe D-brane is entirely outside the black hole horizon, i.e., the Minkowski embedding.(dark green line).
At higher temperature, the probe D-brane partially fall through the black hole horizon, i.e.,
the black hole embedding.
} \label{config}
}

Motivated by these agreements with QCD, 
other region of the QCD phase diagram, i.e., finite temperature with finite baryon density $n_b$ or
finite chemical potential $\mu_q=\mu_b/N_c$ was investigated in \cite{findends,chemical,chemical8,erdmenger,korea,japan,karch3}, see also \cite{Erdmenger:2007bn}. 
The introduction of a finite baryon density corresponds to
the presence of a U(1) gauge field, which is a diagonal part of the gauge group U(N$_f$), on the probe D-brane.

In \cite{findends}, this direction is explicitly investigated in the D3/D7 system.
The boundary gauge theory is SU($N_c$) $\cal{N}$=4 super Yang-Mills theory with 
$N_f$ $\cal{N}$=2 hypermultiplets in the fundamental representation.
There, physical quantities such as quark condensate in finite $n_b$ continuously change from 
those in $n_b=0$.
One of the most crucial difference between zero and nonzero baryon density systems is in the presence finite baryon density,
Minkowski embeddings are unphysical and do not play any role, while
black hole embeddings 
cover the whole range of the temperature above deconfinement phase transition.
This is different from $n_b=0$ case where black hole embeddings cover only high temperature region and Minkowski embeddings cover 
only low temperature region\cite{Mateos:2006nu,Mateos:2007vn}.
In the overlapping region, there is a phase transitions from a Minkowski embedding to a black hole embedding.

The reason why Minkowski embeddings are unphysical in the presence of finite 
baryon density is as follows;
The electric field on the D-brane represents dissolved fundamental strings.
Since Wess-Zumino coupling is inactive in this brane configuration,
fundamental strings never `leak' from the brane and the local 
baryon number density is still nonzero above the horizon.
Since the fundamental strings cannot terminate,
we have to have fundamental strings stretching from 
the probe D-brane to the horizon.
At this junction point, the tension of the probe D-brane is always smaller than that of the fundamental
strings.
That means that there is no force balanced configurations in the Minkowski embeddings 
at finite baryon density and the brane must fall down through the horizon.
See \cite{findends, Mateos:2007vc} for more detail discussions.
Instead 
at very low temperature a black hole embedding mimics a Minkowski embedding, i.e.,
a very long and narrow spike, which corresponds to a bundle of dissolved fundamental strings
on the probe D-brane, stretches down to the horizon. 
For smaller $n_b$, there is a first order phase transition
from a black hole to another black hole embedding,
which is very similar to a Minkowski to a black hole embedding in the zero baryon 
density case.
Above the critical density $n^* _{b}$, there is no phase transition anymore.

However it was shown in \cite{findends} that there is an unstable region near
the phase transition line.
It is expected that this unstable region finally decays to
other stable state. 
However since this region minimizes the free energy, there is no state to decay and 
the final state was missing.

This puzzle was recently addressed in \cite{Mateos:2007vc}.
In the grand canonical ensemble, we fix not a baryon density but a chemical potential.
Since a Minkowski embedding can have a finite chemical potential without a baryon density,
this embedding is physical in the grand canonical ensemble and plays an important role.
The black hole embeddings cover only larger chemical potential or higher temperature region.
On the other hand, the Minkowski embeddings cover the whole value of $\mu$ with
temperature lower than $T_{fun}$ (see fig.2 in \cite{Mateos:2007vc}).
There are phase transitions inside the overlapping region from Minkowski 
to black hole embeddings.
One of the most remarkable results is that the unstable region is
thermodynamically unfavourable in the grand canonical ensemble.
Since in the thermodynamical limit, the grand canonical and the canonical
ensemble should give the same answer\cite{thermo}, the unstable region is
not a true ground state but it should be replaced by an
 inhomogeneous phase of a stable black hole embedding and a Minkowski embedding in the canonical ensemble.

The aim of this paper is to investigate the generality of the phase structure in Dp/Dq system.
We analyze the D4/D6 system.
Physical properties such as 
the phase transitions from a black hole to a black hole embedding in the canonical ensemble and 
from a Minkowski to a black hole embedding in the grand canonical ensemble and the existence of the unstable region are the same. 
However the inhomogeneous phase has a more complicated and interesting structure.
There is a new additional phase transition.
Figure \ref{muT} shows the phase diagram of the D4/D6 system.
As in the case of the D3/D7 system, the black hole embeddings cover only 
high temperature or high chemical potential region. The boundary
of the black hole embeddings is plotted in the green line.
In the region surrounded by the blue line, there are more than one value of $n_q$
for each $\mu_q/M_q$ and $T/\bar{M}$.
The red line is the line of the phase transitions.
On the scale of (a), the difference between the red and the green line 
is very subtle.
An interesting structure is found in (b) and (c).
Near $\mu_q/M_q=0.16$, $T/\bar{M}=0.77$, the red line
separates into two branches. 
The upper branch represents phase transitions from black hole to black hole embeddings.
In the field theory side, this corresponds to a discontinuous change of 
meson lifetime.
Below this phase transition temperature, the spectral function of mesons would show 
sharp peaks representing longer lifetime quasiparticles, while above the phase transition,
the peaks would be flatter and the lifetime of the quasiparticles would be shorter.
The lower branch represents phase transitions from Minkowski to black hole embeddings.
We will discuss the relation between these phases and the unstable region in Section 3.

%

%
\FIGURE{
\begin{tabular}{cc}
  \includegraphics[width=0.5 \textwidth]{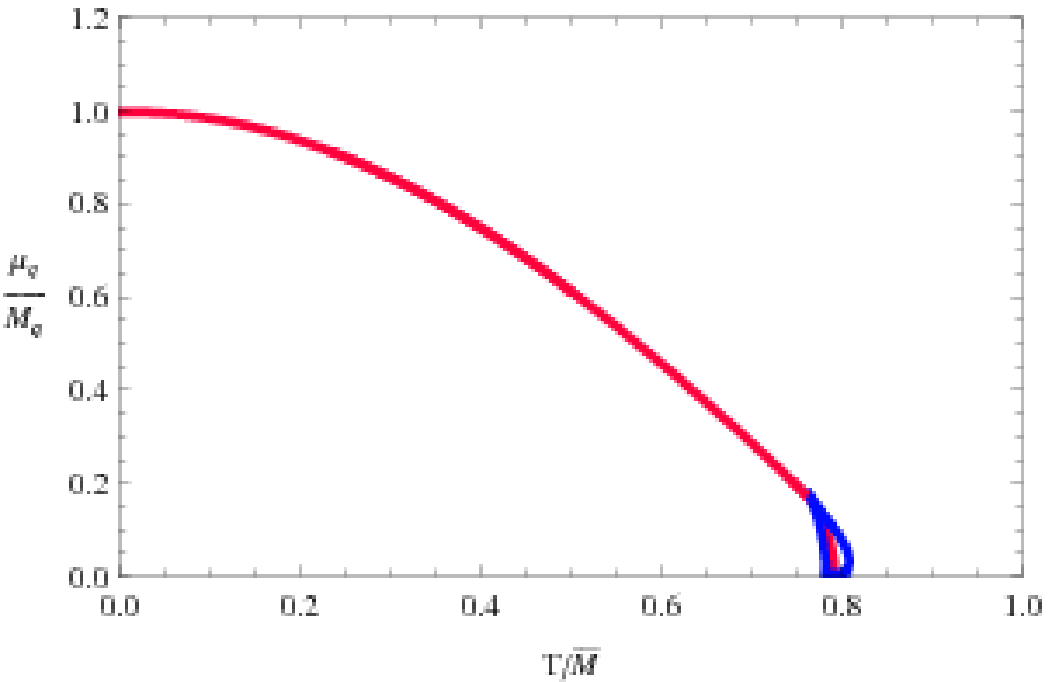}
 \put(-150,60){$\nq=0$}
 \put(-70,100){$\nq \neq 0$} &
  \includegraphics[width=0.5 \textwidth]{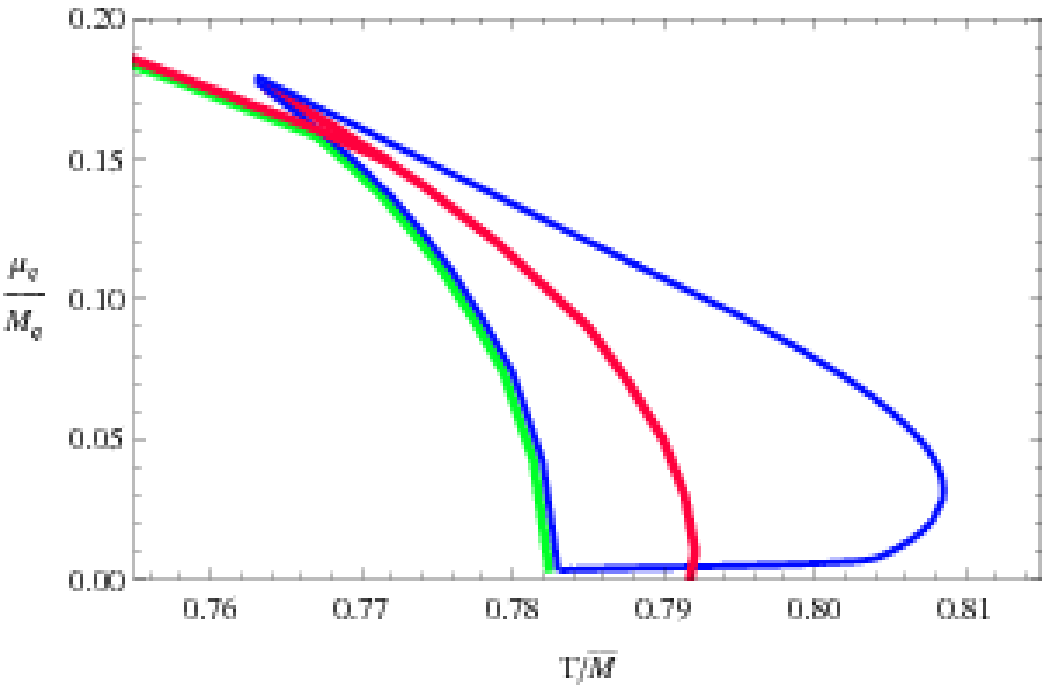}\\
~~~~~~(a) &~~~~~~~~ (b)\\
  \includegraphics[width=0.5 \textwidth]{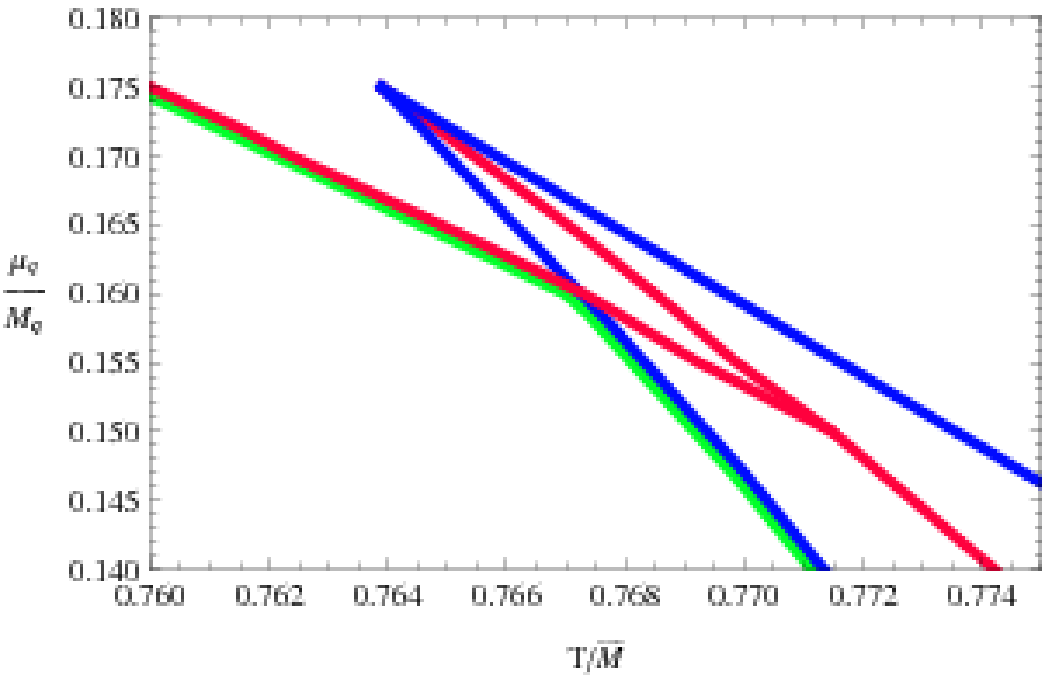}&\\
~~~~~~~~~(c)
\end{tabular}
\caption{Phase diagram: Quark chemical potential $\mu_q/M_q$ versus temperature $T/\bar{M}$. 
The red line separates the phase of Minkowski embeddings (small temperatures, small $\mu_q/M_q$)
from black hole embeddings. Figure (b) zooms in on the region near the end of this line and also depicts the boundary of the region accessed by the black hole embeddings (green) and a small region 
(enclosed by the blue curve) where more than one black hole embedding in available for a given 
value of $\mu_q$ and $T$.
Figure (c) zooms in on the region near the branches of the red lines. 
The two branches of the red lines show the
phase transitions, the lower one is from the Minkowski embeddings to the black hole embeddings and the upper
 one is from the black hole embeddings to the black hole embeddings. 
} 
\label{muT}
}
%



\section{Holographic framework}

In this section we explain the D4/D6 system.
Closed string modes on the near horizon geometry of D4 brane is dual to maximally supersymmetric five dimensional super Yang-Mills
theory with large $N_c$ and at strong coupling on the boundary .  In order to obtain four dimensional gauge theory at low energy scale, we compactify
one space direction, say $x_4$, with a radius $M^{-1} _{KK}$.
For energy scale $E<< M_{KK}$, the boundary gauge theory is effectively four dimensional.
Since we want to study the thermal properties of fundamental fields coupling to Yang-Mills field,
we choose the supersymmetry-breaking spin structure, i.e., the anti-periodic boundary condition for fermions in
 $x_4$ direction.
The fermions in the vector multiplet get masses of order $M_{KK}$ at tree level
and the scalars possibly get masses of order $g^2 _4N_{c}M_{KK}$ at one loop level.
So the gauge theory is effectively non-supersymmetric four dimensional Yang-Mills theory at low energy scale.\footnote{
However the Kaluza-Klein modes do not decouple within the supergravity approximation\cite{Witten:1998zw}.
For example, the lightest glueball spectrum is of the same order as the strong coupling scale\cite{COOT98}.
}
By fixing the boundary geometry, this system undergoes a phase transition
from a thermal AdS phase to an AdS black hole phase at certain temperature $T_{dec}$.
In the dual language, this transition is deconfinement/confinement phase transition\cite{Witten:1998zw}.
This system describes only adjoint matters.
To approach QCD in this holographic framework, quarks or fundamental matter must be included.
The introduction of fundamental matter in the holographic framework is 
demonstrated in \cite{flavour}.
According to their work, a probe D-brane plays a role of fundamental matter in the dual picture.
In our case, inserting N$_f$ D6 branes into this black-D4 background corresponds to the coupling of
$N_f$ hypermultiplets in the fundamental representation to the Yang-Mills fields with the gauge group SU$(N_c)$. The quarks arise from the lightest mode of the fundamental strings connecting between D4 and D6 branes. In the decoupling limit, the mesons are dual to the fundamental strings whose both ends
are attaching on D6 branes.  This means that the fluctuations on the D6 branes describe the mesons
in the dual gauge theory.

\subsection{Black D4 background}
The supergravity solution corresponding to the decoupling limit of $N_c$ coincident 
D4-branes in the black hole phase is in the string frame 

\bea
ds^2&=&{1\over 2}\left({\varrho \over L}\right)^{3/2}\left[-{f^2\over \tilf}dt^2+\tilf dx_4 ^2\right]+\left({L\over \varrho}\right)^{3/2}
{\tilf ^{1/3}\over 2^{1/3}}\left[d\varrho^2+\varrho^2d\Omega_4 ^2\right]\\
e^{\Phi}&=&\left({u\over L}\right)^{3/4}=\left({\varrho ^3\over 4L^3}\right)^{1/4}
\tilde{f}^{1/2}
\label{bgmet}
\eea
where $f=1-{u_{0} ^3\over \varrho^3}$ and $\tilf=1+{u_0 ^3\over \varrho^3}$
\footnote{This metric (isotropic coordinate) is related to the ordinary coordinate
as $(\un\rho)^{3/2}-u^{3/2}=\sqrt{u^3-\un^3}$
}.


$u_0$ is the location of the horizon and the AdS radius $L$ is given by
\beq
L^3=g_s N_c \pi l_s ^3.
\eeq
Hawking radiation appears in this background with the temperature given by the surface gravity 
$T=\kappa/2\pi$, or by the regularity of the Euclidean section, 
\beq
T={3\over4\pi}\sqrt{{u_0\over L^3}}.
\eeq
This temperature is identified with the temperature of the boundary gauge theory.
The dual boundary gauge theory is maximally supersymmetric Yang-Mills theory on R$^{1,4}$
with coordinates $\{t,\vec{x} _{4}\}$.

As we explained above, we compactify on a circle with radius $1/M_{KK}$
with supersymmetry breaking boundary condition.
At low energy $E<<M_{KK}$ the effective field theory is four dimensional non-supersymmetric 
Yang-Mills theory.
The string coupling constant in the bulk theory and the gauge coupling constant in
the dual boundary theory is related through
\beq
{g^2 _{5}}=(2\pi)^2g_{s}l_{s},~~~~ g^2 _{4}=g^2 _{5}M_{KK}
\eeq 
where $g_{s}=e^{\phi_{\infty}}$ and $g_{4,5}$ are the Yang-Mills coupling constants in four and five dimensions.

\subsection{D6-brane embeddings}

We  introduce $N_f$
D6-branes in this background in the following configuration;

\begin{center}
\begin{tabular}{ccccccccccc} \hline
  &0 &1 &2 &3 &4 &5 &6 &7 &8 &9   \\\hline
D4&$\times$&$\times$&$\times$&$\times$&$\times$&&&&& \\\hline
D6&$\times$&$\times$&$\times$&$\times$&&$\times$&$\times$&$\times$&&\\\hline
\end{tabular}
\end{center}

We split the AdS radial direction and $S^4$ into two parts so that 
$SO(3)\times SO(2)$ symmetry is manifest.
We define the radial coordinate $r$ in ($x^5,x^6,x^7$) directions and 
$R$ in ($x^8,x^9$) directions where
$\rho=\varrho/\un$ is a dimensionless coordinate.
\beq
\rho^2=r^2+R^2,~~~r=\rho\sin{\theta},~~~R=\rho\cos{\theta},
\eeq
and
\bea
d\rho^2+\rho^2d\Omega_{4} ^2&=&d\rho^2+\rho^2(d\theta^2+\sin^2{\theta}d\Omega_2 ^2+\cos^2{\theta}d\phi^2)\\
&=&dr^2+r^2d\Omega_2 ^{2}+dR^2+R^2d\phi^2
\eea
The probe D6 brane embedding is parametrized by a function,
$\chi(\varrho)\equiv\cos\theta(\varrho)$.
The induced D6 metric is
\beq
ds^2={1\over2}\left({\varrho\over L}\right)^{3/2}\left[-{f^2\over \tilf}dt^2+\tilf dx_3 ^2\right]+
\left({L\over \varrho}\right)^{3/2}{\tilf ^{1/3}\over 2^{1/3}}\left[\left(1+{\varrho^2\dot{\chi}^2\over 1-\chi^2}\right)d\varrho^2+
\varrho^2\left(1-\chi^2\right)d\Omega_2 ^2 \right]
\eeq

Now we introduce U(1) gauge field $A(\varrho)dt$ on the probe D6 branes.
The D6 brane action per unit spacetime volume of the gauge theory is
\bea
I_{D6}&=&-N_{f}T_{6}\int e^{-\Phi}\sqrt{-\det(P[G]_{ab}+2\pi\alpha'F_{ab})}\\
&=& -N_{f}T_{6}{\Omega_2 \over 4}  \int
d\varrho  \tilf^{4/3}\varrho^2\left(1-\chi^2\right)\sqrt{f^2\tilf^{-2/3}\left(1+{\varrho^2\dot{\chi}^2\over 1-\chi^2}\right)
-k\dot{A}^2}
\label{actionS}
\eea
where 
\beq
k=(2^{2/3}2\pi l_{s}^2)^2, 
\eeq
and the D6 brane tension is
\beq
T_{6}={2\pi\over g_{s}(2\pi l_{s})^7}.
\eeq


Let us analyze the behaviour of the electric field $A$.
The electric displacement $d$ is defined by 
\beq
d={N_{f}T_6 \Omega_2 \over 4}{\tilf^{4/3}\varrho^2(1-\chi^2)k\dot{A}\over \sqrt{f^2\tilf^{-2/3}(1+{\varrho^2\dot{\chi}^2
\over 1-\chi^2})-k\dot{A}^2}}
\eeq

The electric displacement represents the number density of dissolved fundamental strings
in the D6 branes.
With the similar analysis to \cite{findends, Mateos:2007vc}
the electric displacement identified with the string number density $n_q$: $d=n_q$.
Since the action does not explicitly depend on $A$,
the equation of motion for $A$ is $d=$constant, or

\bea
\dot{A}&=&{df\tilde{f}^{-{1\over 3}}\sqrt{1+{\varrho ^2\dot{\chi}^2 \over 1-\chi ^2}} \over
\sqrt{d^2k+\left({N_{f}T_6 \Omega_2  k \over 4}\right)^2 \tilde{f}^{8/3}\varrho^4\left(1-\chi^2\right)^2 }  }\\
\eea
Asymptotically $\chi$ goes to zero and $f$ and $\tilf$ go to one.
The leading order of the equation is then
\beq
\dot{A}={d \over {N_{f}T_6 \Omega_2  k \over 4} \varrho^2}+{\cal O}\left({1\over \rho^3}\right)
\eeq
So the asymptotic behaviour of the gauge field is
\bea
A\sim\mu_q-{4d\over N_{f}T_6 \Omega_2 k}{1\over \varrho} + \cdots 
\label{aasym}
\eea
where $\mu$ is
\beq
\mu_q=d\int^{\infty} _{\un}d\varrho 
{f\tilde{f}^{-{1\over 3}}\sqrt{1+{\varrho ^2\dot{\chi}^2 \over 1-\chi ^2}} \over
\sqrt{d^2k+\left({N_{f}T_6 \Omega_2  k \over 4}\right)^2 \tilde{f}^{8/3}\varrho^4\left(1-\chi^2\right)^2 }  }.
\label{muqd}
\eeq

Here we set $A(u_{0})=0$. 
This condition comes from the regularity of the one form at the horizon.
The horizon contains a bifurcation surface where a Killing vector $\partial_{t}$ vanishes. 
In order for the gauge field, which is a one form $A_{t}dt$, to be well defined,
the component $A_{t}$ must vanish there. See also \cite{findends,Mateos:2007vn}.

According to the holographic dictionary, the dual operator is schematically
\beq
{\cal O}_q=\psi^{\dagger}\psi+q^{\dagger}{\cal D}_{t}q
\label{adualope}
\eeq 
where $\psi$ and $q$ are left and right Weyl fermions and scalar fields (see also \cite{findends}). 
From \reef{aasym} and \reef{adualope}, we identify $\mu_q$ with a quark chemical potential.

For convenience, we also use dimensionless quantities and a gauge field,
\beq
\tilde{d}={4\over N_{f}T_6 \Omega_2  \sqrt{k}}{d\over u_0 ^2}, ~~~~\tilde{\mu}={\sqrt{k}\over \un}\mu 
,~~~~\tilde{A}={\sqrt{k}\over \un}A.
\label{ddtil}
\eeq
The equation of motion for $\chi$ is
\bea
{d\over d\varrho}&&
\left(f\tilf{\varrho^4\dot{\chi}\over \sqrt{1+
{\varrho^2\dot{\chi}^2\over 1-\chi^2}
-{k\dot{A}^2\over f^2\tilf^{-2/3}}
}}\right)\\
&&+
\left(f\tilf{\varrho^2\chi \over \sqrt{1+
{\varrho^2\dot{\chi}^2\over 1-\chi^2}
-{k\dot{A}^2\over f^2\tilf^{-2/3}}
}}\right)\left(2+{\varrho^2\dot{\chi}^2\over 1-\chi ^2}
-2{k\dot{A}^2\over f^2\tilf^{-2/3}}\right)=0
\label{eomchi}
\eea
The boundary conditions for $\chi(\rho)$ at the horizon is determined by the regularity 
and it gives $\chi\bigr|_{\rho=1}=\chi_0$ and $d\chi/d\rho\bigr|_{\rho=1}=0$ for $0\leq\chi<1$.
For Minkowski embeddings it is convenient to use $R,r$ coordinates instead of $\chi,\rho$ coordinates.
The boundary conditions for Minkowski embeddings at $r=0$ are $R\bigr|_{r=0}=R_0$ and $dR/dr\bigr|_{r=0}=0$ for $1<R_0$.
The asymptotic forms ($\rho \rightarrow \infty$) for $\chi$ is
\bea
\chi &\sim {m\over \varrho}+{c\over \varrho^2} + {\cal O}\left({1\over \varrho^3 } \right) \\
&=  {\tilde{m}\over \rho}+ {\tilde{c}\over \rho^2} + {\cal O}\left({1\over \rho^3 } \right) ,
\eea
where we define dimensionless quantities
\beq
\tilde{m}={m \over u_0}={3^2 m \over (4\pi)^2L^3 T^2},~~~
\tilde{c}={c\over u_0 ^2}={3^4 c\over (4\pi)^4 L^6 T^4}.
\eeq
Holography relates these quantities to a quark mass and a condensate\cite{Kruczenski:2003uq} by
\bea
M_{q}&=&{u_0 \tilde{m}\over 2^{5\over3}\pi l^2 _{s}}
\\
<{\cal O}_{m}>&=&-2^{5/3}\pi^2 l^2 _{s}N_{f}T_{D6}u^2 _{0}\tilde{c}
\eea
A bare quark mass $M_{q}$ is an asymptotic distance between D4 and D6 brane in flat space.
The operator ${\cal O}_m$ is the variation of the mass term in the microscopic Lagrangian,
i.e., ${\cal O}_{m}=-\partial_{M_q}{\cal L}$, and the schematic form is
\beq
{\cal O}_{m}=\bar{\psi}\psi+q^{\dagger}\Phi q+M_{q}q^{\dagger}q,
\eeq
where $\Phi$ is one of the adjoint scalars. See also \cite{findends,Mateos:2007vn}.
We assume that the vacuum expectation value of the fundamental scalar fields q vanishes because whenever the scalar fields acquire the nonzero expectation value, the energy density increases\cite{Kruczenski:2003uq}.
In that case, $<{\cal O}_{m}>$ is equal to $<\bar{\psi}\psi>$ and represents a quark condensation.
Given the above result, we show various figures in terms of $T/\bar{M}\equiv 1/\sqrt{\tilde{m}}$.

Let us study more on the chemical potential.
Once the equation of motion for $\chi(\varrho)$ \reef{eomchi} is solved, the chemical potential 
is obtained from \reef{muqd}.
In general, we have to resort to numerical calculation.
However, we can extract analytic properties in the limiting cases of small and high temperatures.
First we consider low temperature or large bare quark mass limit $T/\bar{M}\rightarrow 0$.
As shown in Figure \ref{config}, at very low temperature
the probe D-brane goes up straight from the horizon $u_0$ to $\sim m$.
The main contribution to the distance between the D4 and D6 brane, or $T/\bar{M}$, comes from this part.
Under this approximation,
\bea
\mu_q&=&d\int^{\infty} _{\un}d\varrho 
{f\tilde{f}^{-{1\over 3}}\sqrt{1+{\varrho ^2\dot{\chi}^2 \over 1-\chi ^2}} \over
\sqrt{d^2k+\left({N_{f}T_6 \Omega_2  k \over 4}\right)^2 \tilde{f}^{8/3}\varrho^4\left(1-\chi^2\right)^2 }  }\\
&\simeq&
d\int^{m} _{\un}d\varrho 
{f\tilde{f}^{-{1\over 3}} \over
\sqrt{d^2k}}\\
&\simeq&m/\sqrt{k}=M_q
\eea
Regardless of $d$, $\mu_q$ goes to $M_q$. This is consistent with Figure \ref{muvsTvarid} where all curves with different $\tilde{d}$ meet on the vertical axis at $\mu_q/M_q=1$.
\FIGURE{
  \includegraphics[width=0.8 \textwidth]{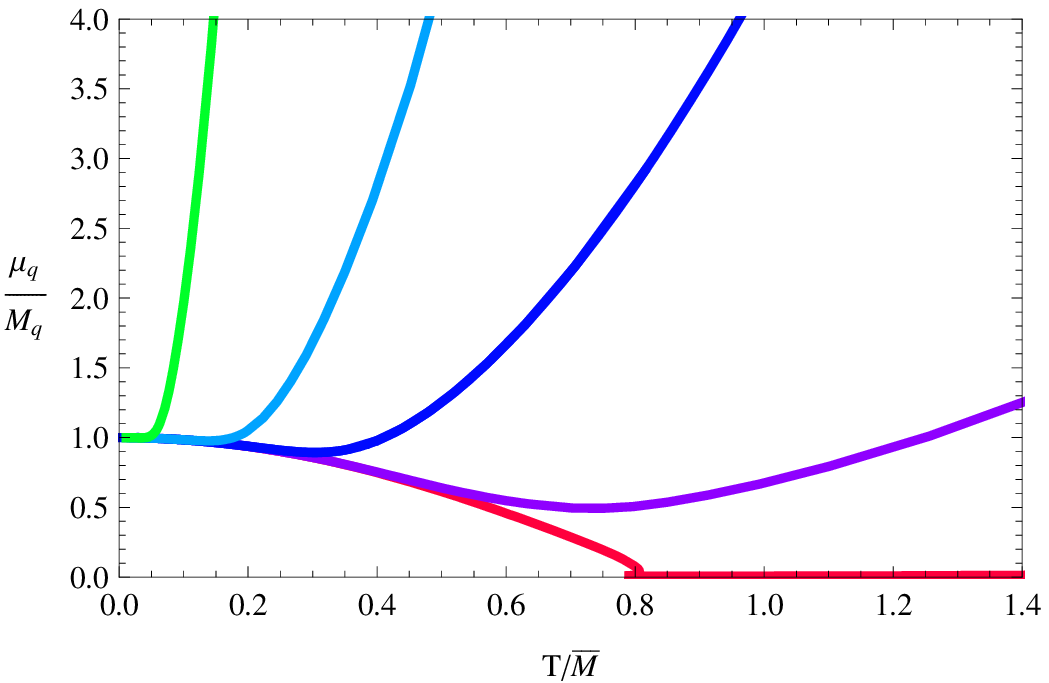}
\caption{Chemical potential $\mu_q/M_q$ versus $T/\bar{M}$ for various values of $\tilde{d}$,
increasing from bottom up:
$\tilde{d}=10^{-2}, 1, 10, 100, 10^4$.
} \label{muvsTvarid}
}
On the opposite limit, i.e.,  high temperature or small bare quark mass limit $T/\bar{M}\rightarrow \infty$, the probe D6 branes are almost flat and intersect the horizon at the
equator. Plugging $\chi \simeq 0$ and $u_0>>1$ into the equation, the leading contribution is
\bea
\mu_q&\simeq&{4d\over \left(N_{f}T_6 \Omega_2  k \right)}{1\over 2^{2/3}u_0}+{\cal O}(d^3)\\
&=&{9 \over 4N_fN_c}{n_q\over T^2}+{\cal O}(n^3 _q)
\eea
There are some notable points.
The first point is that for any value of $n_b$, $\mu_q$ goes to zero as temperature goes to infinity. 
In this sense, Figure \ref{muvsTvarid} might be misleading.
Each line in the figure shows $\mu_q/M_q$ versus $T/\bar{M}$ for fixed $\tilde{d}$ not $d$.
In each line, higher temperature corresponds to a higher baryon number density and 
lower temperature corresponds to a lower baryon number density.
The second point is for fixed $\mu_q$, $n_b$ changes as $\sim T^2$.
Since in the grand canonical ensemble we survey fix $\mu_q$ planes,
this equation suggests that the baryon density monotonically increases at very high temperature
for any fixed $\mu_q/M_q$.

\section{Thermodynamics}

We move on to the thermodynamics of the D6 brane, or the fundamental matter.
As we will see the canonical ensemble is not a suitable one for studying the 
phase diagram since it includes an unstable region.
Our main focus here is the grand canonical ensemble.
A similar unstable region can also be found in the D3/D7 system \cite{Mateos:2007vc}. 
We will briefly mention on the canonical ensemble.

Euclidean path integral of D6 branes give thermal partition function\cite{hawk}.
Since the classical solution of the equation of motion is the saddle point of the path integral,
the on-shell action gives main contribution to the Gibbs free energy $W$, i.e., $W=TI_{E}$.
From the grand canonical point of view, a brane configuration which minimizes the 
Gibbs free energy is thermodynamically favourable.
However,  naive calculation of thermal quantities include divergences from IR region.
For example, we can clearly see that the action diverges as
\beq
\int d\rho \rho^2 \simeq {\rho^3 _{max}\over3}.
\eeq

In order to obtain physically meaningful, finite quantities, we need to renormalize them.
The renormalization procedure of probe D-brane is studied in \cite{karch1}.
According to that, renormalization is done by introducing boundary terms 
for probe branes on the cut-off plane so that the 
divergences cancel. For the D4/D6 case, this boundary term is concretely 
studied in \cite{Mateos:2007vn}.
Before starting the detail analysis, we define a normalization constant
\beq
{\cal N}={\pi\over T}N_{f}T_{D6}u^3 _{0}.
\eeq
%
%
%
Inserting the asymptotic expansion of $\chi$ and $\dot{A}$ into \reef{actionS}, the action is
\beq
{I_{reg}\over {\cal N}} = \int d\rho \rho^2\left(1-{\tilde{m}^2\over \rho^2}-{2\tilde{m}\tilde{c}\over\rho^2}\right)\left(1+{\tilde{m}^2\over 2\rho^2}\cdots -{\tilde{d}^2\over 2\rho^4}\right).
\eeq
The contribution to the action from the gauge field is the term proportional to $\tilde{d}^2$.
This gives no new divergence.
So we can use the same boundary term as that of $\tilde{d}$=0 case to renormalize the divergence.
We apply the following boundary term per unit spacetime volume of the gauge field \cite{Mateos:2007vn},
\bea
I_{bound}&=&-{\Omega_2\over 3}L^3 T_{6}N_{f} \sqrt{\gamma}\left(1-{3\over 2}\chi^2\right)\mid_{\rho=\rho_{max}}\\
&=&-{u_0 ^3\Omega_2 T_{6}N_{f} \over 12} \left(\rho_{max} ^3-{3\over2}\tilde{m}^2 \rho_{max} -3\tilde{m}\tilde{c}\right),
\eea
where $\gamma$ is the induced metric at $\rho=\rho_{max}$
\beq
ds_{\gamma}^2={1\over 2}\left({u_0\rho_{max}\over L}\right)^{3/2}\left(-{f(\rho_{max})^2\over \tilf\left(\rho_{max}\right)}dt^2+\tilf\left(\rho_{max}\right) dx_3 ^2\right).
\eeq
The total action is then,
\bea
{I_{tot}\over {\cal N}}&=&{I_{reg}\over {\cal N}}+{I_{bound}\over {\cal N}}\\
&=&\left[ G(\tilde{m},\dot{\tilde{A}})-{1\over 3}\left(\rho_{min} ^3-{3\over2}\tilde{m}^2 \rho_{min} -3\tilde{m}\tilde{c}\right)\right],
\eea
where $G(\tilde{m},\dot{\tilde{A}})$ is
\beq
G(\tilde{m},\dot{\tilde{A}})=\int _{\rho_{min}} ^{\infty} d\rho\left( \tilf^{4/3}\rho^2\left(1-\chi^2\right)\sqrt{f^2\tilf^{-2/3}\left(1+{\rho^2\dot{\chi}^2\over 1-\chi^2}\right)
-\dot{\tilde{A}}^2}-\rho^2+{\tilde{m}^2\over 2}\right)
\eeq
From the thermodynamical point of view, this action is identified the Gibbs free energy $W(T,\mu_{q})$ via $W=TI_{E}$.

In the canonical ensemble with fixed $n_q$, we use the Helmholtz free energy. 
Similar to \cite{findends, chargeblack}, the Helmholtz free energy is associated with the 
Legendre transform of $I_{E}$. 
\bea
{\tilde{I}_{E}\over{\cal N}}={{I}_{D6}\over{\cal N}}+{\int   d \dot{A}\over {\cal N}}
\eea
which is function of the temperature and the baryon density.
We identify $F(T,n_q)=TI_{E}$ where $F(T,n_q)$ is the Helmholtz free energy.
Since there is no contribution from the electric displacement to the boundary term 
, the Legendre transformed action is
\beq
{\tilde{I}_{E}\over{\cal N}}= H(\tilde{m},\tilde{d}) -{1\over 3}\left(\rho_{min} ^3-{3\over2}\tilde{m}^2 \rho_{min} -3\tilde{m}\tilde{c}\right)
\eeq
where $H(\tilde{m},\tilde{d})$ is
\beq
H(\tilde{m},\tilde{d})=
\int _{\rho_{min}} ^{\infty} d\rho\left[
f\tilf^{-1/3}\sqrt{1+{{\rho^2\dot{\chi}^2}\over 1-\chi^2}}\sqrt{\tilf^{8/3}\rho^4(1-\chi^2)^2+\tilde{d}^2}
-\rho^2+{\tilde{m}^2\over 2}\right].
\eeq
%
%
%
%
%
We evaluated the free energy numerically. The qualitative feature is very similar to
that of the D3/D7 system (see Figure 9 and 10 in \cite{findends}).
For smaller value of $\tilde{d}/\tilde{m}^2$, there is a line of first order phase transition
from black hole to black hole embeddings and above a critical value $\left(\tilde{d}/\tilde{m}^2\right)^*$ there is no phase transition.
Note that since $\tilde{d}$ depends on a temperature $n_q/T^4$, 
we use a temperature independent quantity $\tilde{d}/\tilde{m}^2 \sim n_q/M^2 _q$ as a parameter.
The blue line in Figure \ref{BhMkPhaseTrans} shows the phase transition line in the canonical ensemble.
The critical temperature and baryon density is
$T/\bar{M}=0.7786$ and $\tilde{d}/\tilde{m}^2=0.041$.

%
%







\subsection{Phase transitions in the grand canonical ensemble}

As we mentioned in Section 1, the black hole embeddings cover only above the green line 
in Figure \ref{muT} while the Minkowski embeddings cover whole value of $\mu_q/M_q$ 
below $\sim T_{fun}$.
This suggests that there is a line of first order phase transitions
for $\mu_q/M_q<1$ in the grand canonical ensemble.
Thermodynamically favourable embeddings minimize the Gibbs free energy for give temperature and
chemical potential.
In this section, we calculate the free energy for various $\mu_q/M_q$.

\FIGURE{
\begin{tabular}{cc}
  \includegraphics[width=0.5 \textwidth]{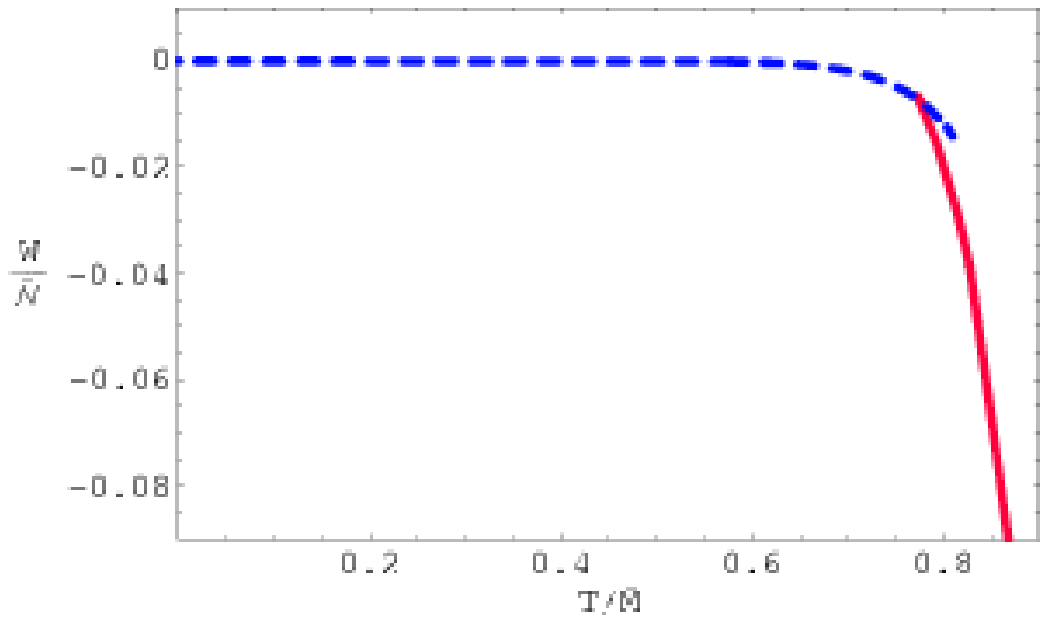}&
  \includegraphics[width=0.5  \textwidth]{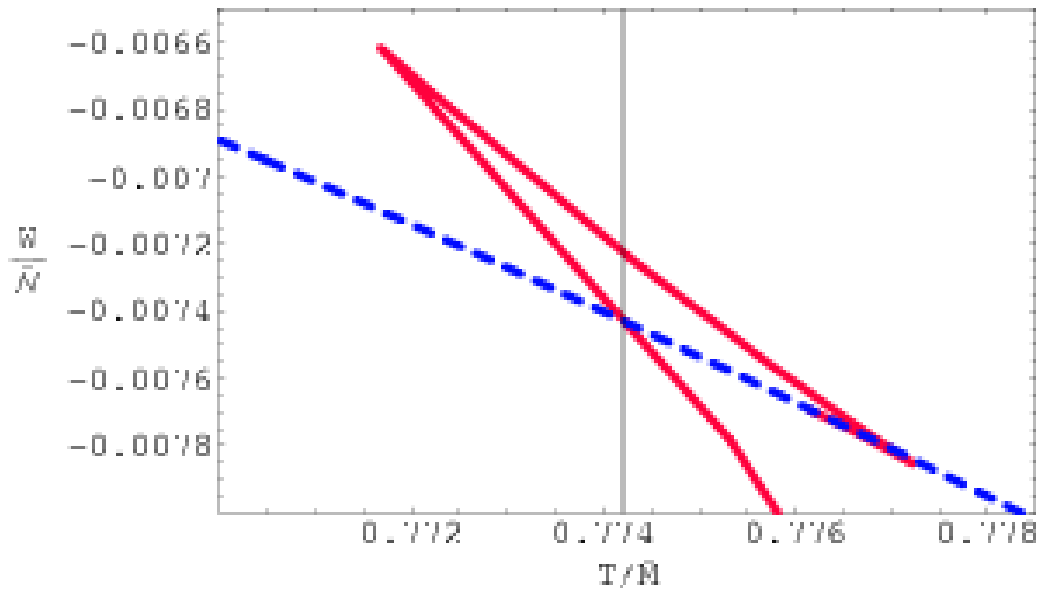}\\
~~~~~~(a)& ~~~~~~(b) 
\end{tabular}
\caption{Free energy versus temperature for $\mu_q/M_q=0.14$.
The blue dotted (red solid) line represents the Minkowski (black hole) branch.
The vertical line marks the temperature of the phase transition.
} \label{WvsTmu014}
}
\FIGURE{
\begin{tabular}{cc}
  \includegraphics[width=0.5 \textwidth]{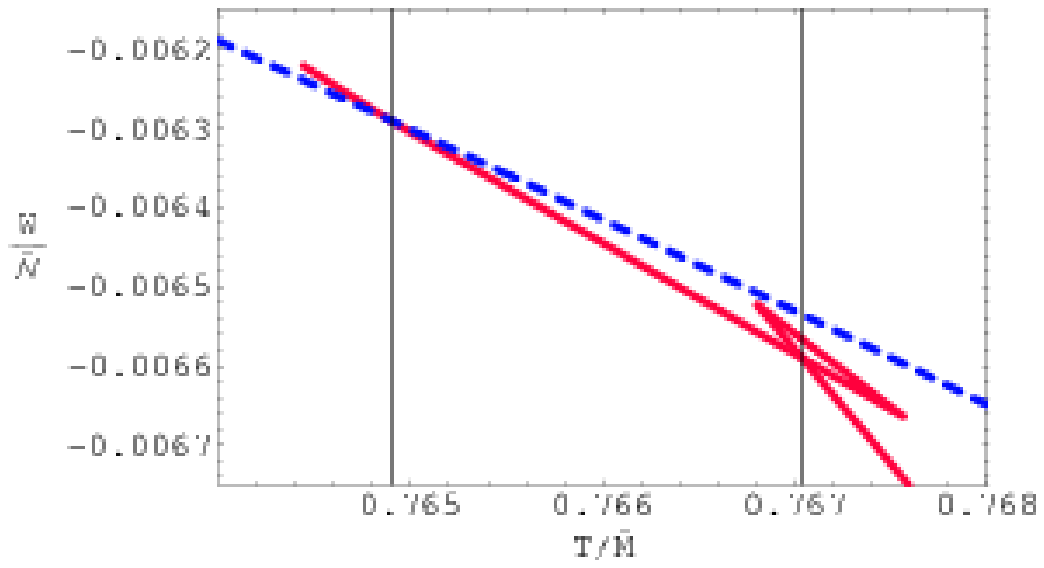}&
  \includegraphics[width=0.5 \textwidth]{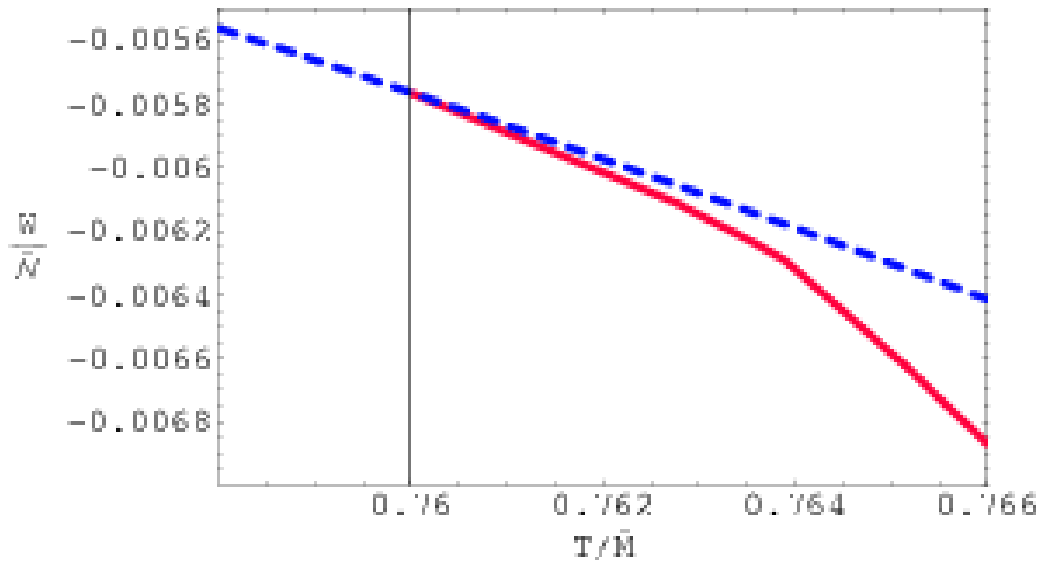}\\
(a)$\mu_q/M_q$=0.165 &(b)$\mu_q/M_q$=0.175
\end{tabular}
\caption{Free energy versus temperature for (a)$\mu_q/M_q$=0.165 and (b)$\mu_q/M_q$=0.175.
(a) shows that there are two phase transitions. One is from the Minkowski embedding to the black hole embedding
at $T/\bar{M}=0.765$ and the other is from the black hole embedding to the black hole embedding
at $T/\bar{M}=0.767$.
(b) shows that there is only one phase transition from the Minkowski embedding to the black hole embedding at $T/\bar{M}=0.76$.
The vertical lines mark the temperature of the phase transitions.
} \label{WvsTmu01650175}
}
\FIGURE{
  \includegraphics[width=0.5 \textwidth]{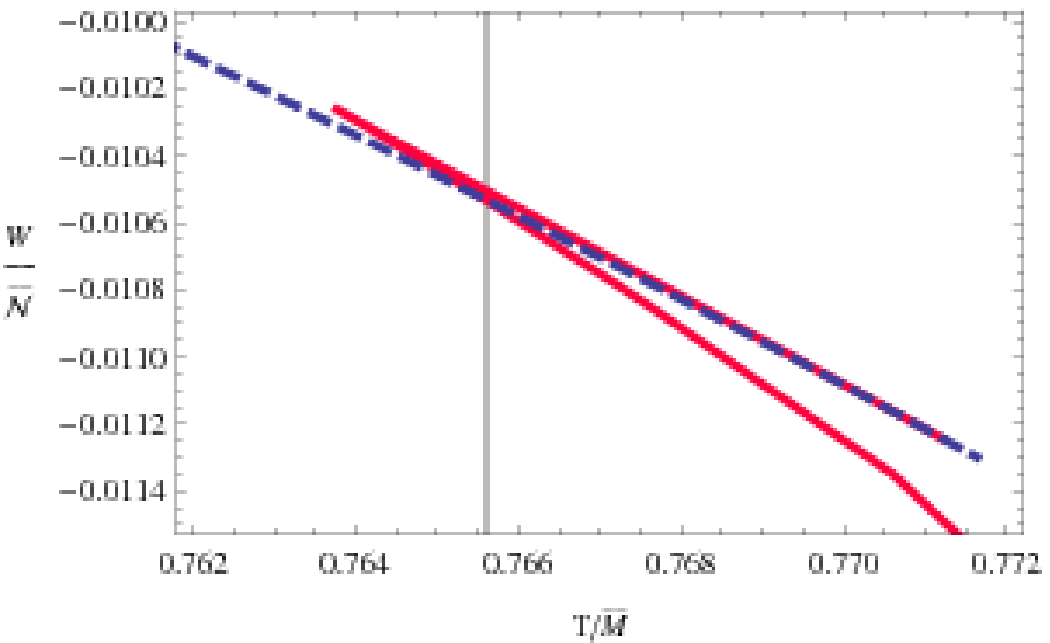}
\caption{Free energy versus temperature for $\mu_q/M_q=0.005$ in the D3/D7 system.
There is only one turnover in the black hole embeddings. This should be compared with
the D4/D6 system (Figure \ref{WvsTmu014}) where there are two turnovers. 
} \label{D3D7WvsTmu0005}
}

Figure \ref{WvsTmu014}(a) shows the Gibbs free energy $W$ normalized by ${\cal \bar{N}}$ versus temperature 
$T/\bar{M}$ for $\mu_q=0.14$ in a broad view
and Figure \ref{WvsTmu014}(b) shows a zoomed in around the phase transition.
Here the normalization is ${\cal \bar{N}}={\cal N}\bar{M}$.
The red line is the black hole embedding and the dashed blue line is the Minkowski embedding.
The starting point of the red line $(T/\bar{M}=0.776$, $W/\bar{{\cal N}}=-0.0077)$ is $\tilde{d}=10^{-3}$. As $\tilde{d}$ becomes larger, the
temperature becomes higher and the free energy becomes smaller, then at $\tilde{d}=5\times10^{-2}$
the temperature starts to decrease and the free energy starts increase.
And at $\tilde{d}=0.17$, the temperature turns to increase and the free energy turns decrease again.
As $\mu$ increases, the starting point of the free energy of the black hole embedding
at very small $\tilde{d}$ goes in the left above direction and
the three fold structure starts to form a swallow tail shape.
At a critical value of $\mu_q/M_q$=0.15, the crossing point of the swallow tail goes down below the
line of the Minkowski free energy.
Above this critical value, there are two phase transitions.
The first one is from a Minkowski embedding to a black hole embedding.
The second one is from a black hole embedding to another black hole embedding.
A representative of this phase is shown in Figure \ref{WvsTmu01650175}(a).
At $\mu_q/M_q=0.165$, there is phase transition from a Minkowski embedding to a black hole embedding 
at $T/\bar{M}=0.765$ and there is a phase transition from a black hole embedding to a black hole embedding at 
$T/\bar{M}=0.767$.
As $\mu$ becomes larger, the swallow tail shrinks smaller and smaller and finally at 
$\mu_q/M_q=0.175$ the black hole to black hole phase transition disappears.
The free energy at this value of $\mu_q/M_q$ is shown in Figure \ref{WvsTmu01650175}(b).
There is a phase transition at $T/\bar{M}=0.76$ and this is the only phase transition in this phase.
These three fold structures in the D4/D6 system
 should be compared to the Figure \ref{D3D7WvsTmu0005} which shows the typical phase transition
for $\mu_q<M_q$ in the D3/D7
system (see also Figure 4 in \cite{Mateos:2007vc}).
The only phase transition is from a Minkowski embedding to a black hole embedding
and there is no phase transition from a black hole to a black hole embedding.

This feature can also be seen from other perspective.
Figure \ref{D4D6dvsTmu016D3D7dvsTmu0005}(a) shows $\tilde{d}/\tilde{m}^2$ versus
$T/\bar{M}$ diagram near the phase transition for $\mu_q/M_q=0.16$. 
As explained above, below a certain temperature there are only Minkowski embeddings.
Around $T/\bar{M}=0.7668$ black hole embeddings appear.
However the Minkowski embeddings are still favoured.
At $T/\bar{M}=0.7673$, the free energy of the black hole embedding becomes equal to the Minkowski
embedding and the system jumps from $\tilde{d}/\tilde{m}^2=0$ to $\tilde{d}/\tilde{m}^2=0.0034$ (A)
As temperature increases, the system goes along the red line until it meets another phase transition point(B).
At $T/\bar{M}=0.7685$, $\tilde{d}/\tilde{m}^2=0.0146$ (B), the system jumps to a configuration which has
the same chemical potential but larger value of $\tilde{d}/\tilde{m}^2=0.0708$(D) in a black hole embedding.
Above this temperature, there is no phase transition and as $T/\bar{M}$ becomes larger $\tilde{d}/\tilde{m}^2$ 
becomes larger.

We also show a $\tilde{d}/\tilde{m}^3$ versus $T/\bar{M}$ diagram for $\mu_q/M_q=0.005$ in the D3/D7 system 
in Figure \ref{D4D6dvsTmu016D3D7dvsTmu0005}(b).
As above, below a certain temperature there are only Minkowski embeddings.
At $T/\bar{M}=0.7656$, the free energy of the Minkowski embedding and the black hole embedding coincides
and the first order phase transition occurs from $\tilde{d}/\tilde{m}^3=0$ to $\tilde{d}/\tilde{m}^3=0.00073$(G). 
Clearly $\mu_q/\bar{M}_q$ is a two-valued function of $T/\bar{M}$ while it is a triple-valued function
in the case of the D4/D6 system, and the region corresponding to
(A)-(B) in Figure \ref{D4D6dvsTmu016D3D7dvsTmu0005}(a) is missing.

\FIGURE{
\begin{tabular}{cc}
  \includegraphics[width=0.5 \textwidth]{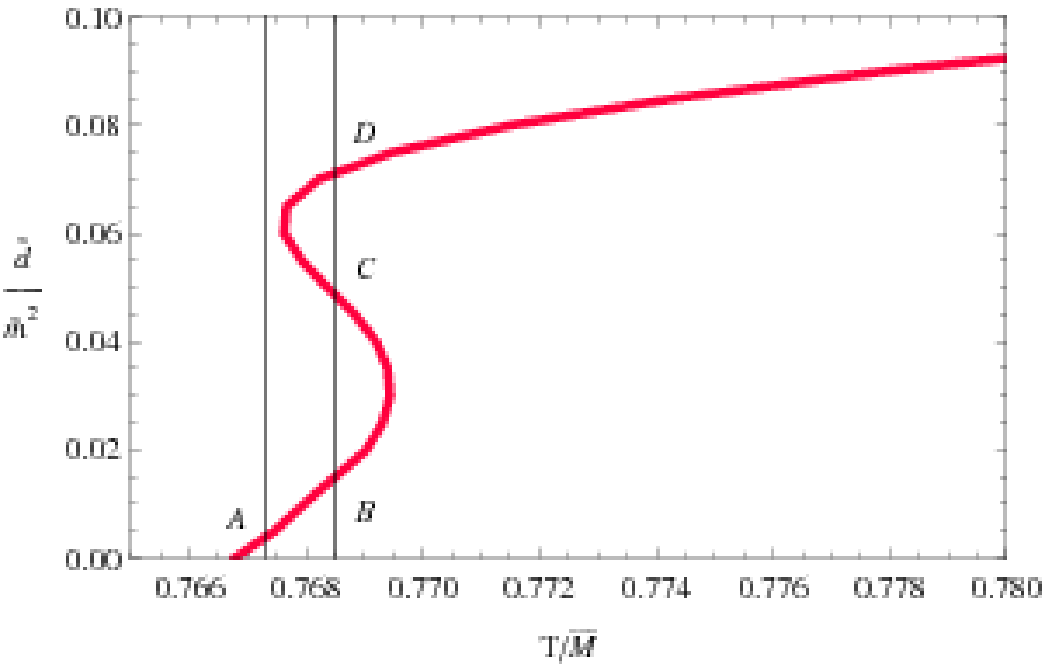}&
  \includegraphics[width=0.52 \textwidth]{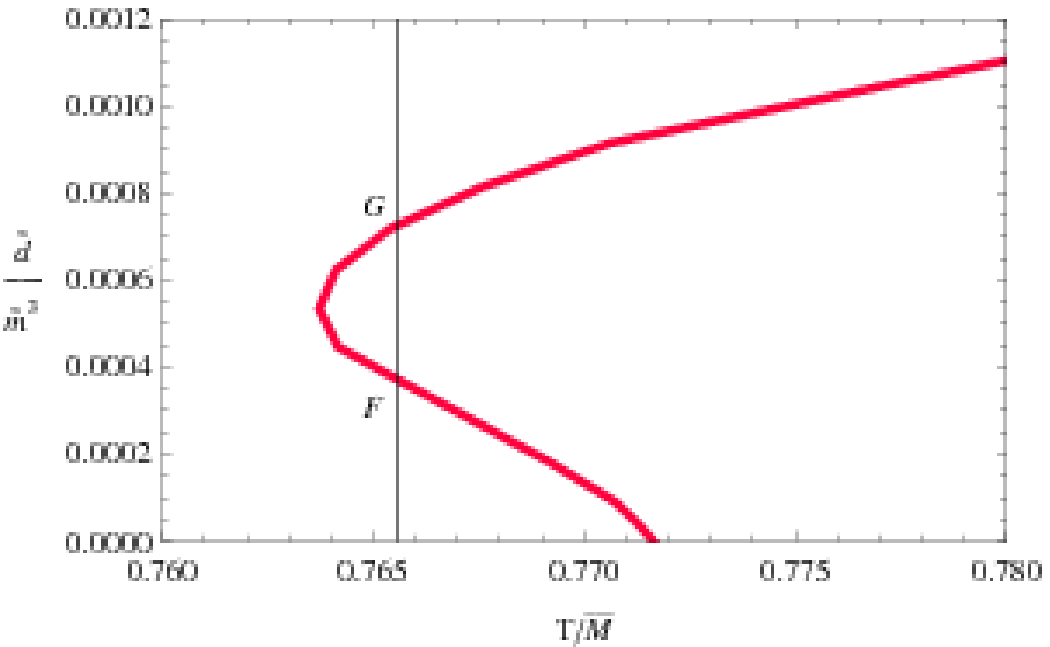}\\
(a) Baryon density versus temperature in D4/D6 &
(b) The same in D3/D7\\
\end{tabular}
\caption{Baryon density versus temperature for (a) $\mu_q/M_q=0.16$ in the D4/D6 system
and (b) $\mu_q/M_q=0.005$ in the D3/D7 system.
In the D4/D6 system, baryon density is a triple-valued function of temperature while 
in the D3/D7 system, baryon density is a double-valued function of temperature. 
 } \label{D4D6dvsTmu016D3D7dvsTmu0005}
}
%


%
\FIGURE{
\rotatebox{270}{
  \includegraphics[width=0.7 \textwidth]{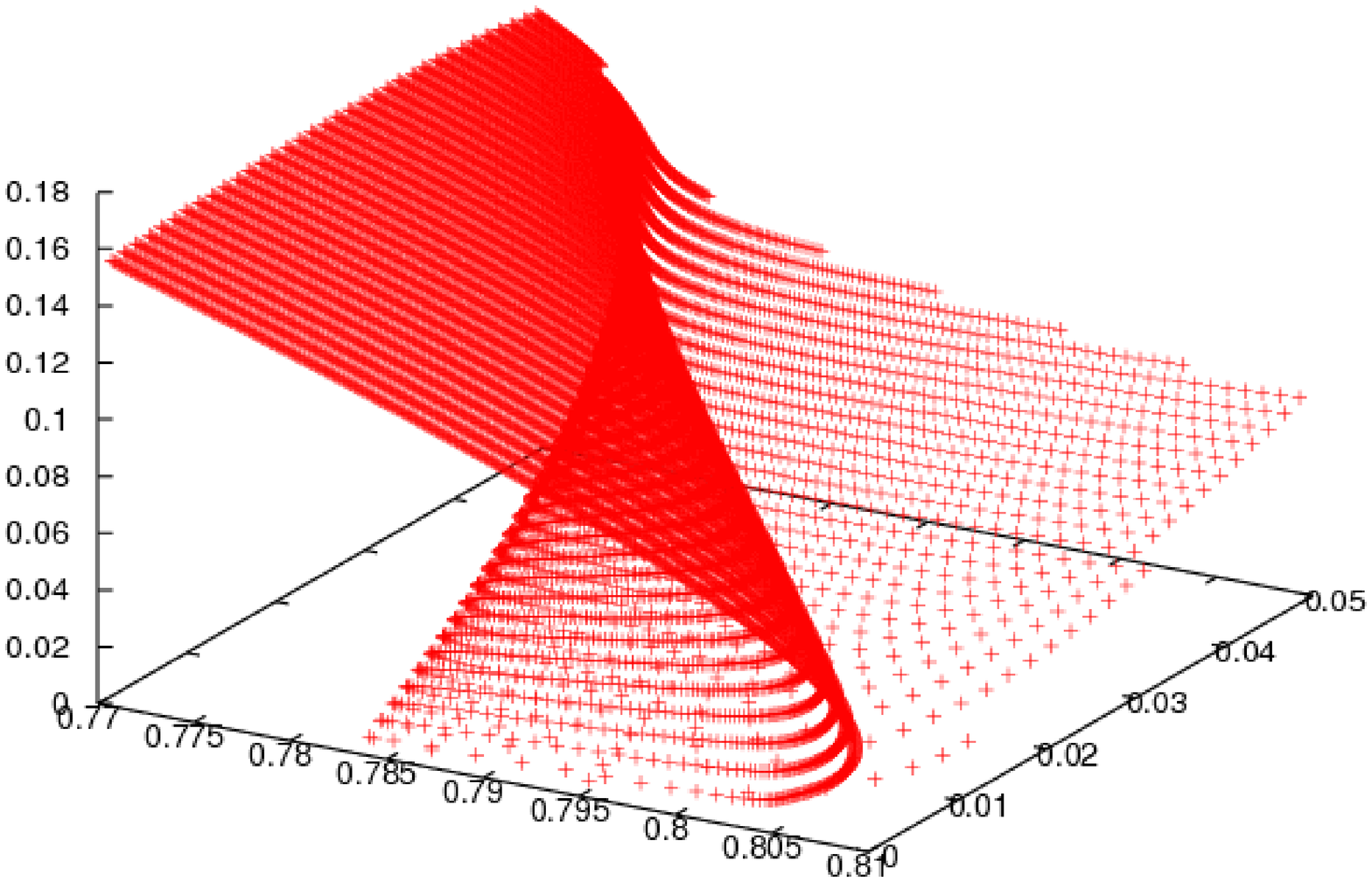}
}
 \put(-430,-160){${\mu_q\over M_q}$}
 \put(-310,-290){${T/\bar{M}}$}
  \put(-100,-260){${\tilde{d}/\tilde{m}^2}$}
\caption{Three-dimensional plot of the chemical potential, the temperature and the charge density
determined by black hole embeddings.}
 \label{3d}}

In the rest of this section, we address another interesting property, i.e., the instability
in the canonical ensemble, and
combine it with the phase transitions discussed above.
The conditions for the stability of the system are given 
\beq
{\partial S\over \partial T}\Bigr|_{\mu_q}>0,~~
{\partial n_{b}\over \partial \mu_q}\Bigr|_{T}>0.
\label{stability}
\eeq
The unstable configuration can be found by examining $\mu_q/M_q$ versus $T/\bar{M}$ and $\tilde{d}/\tilde{m}^2$
diagram(Figure \ref{3d}).
This figure covers only a small range of the surface near the phase transition in the canonical ensemble.
For fixed $n_b$, or $\tilde{d}/\tilde{m}^2$, the system jumps from the top to the bottom of the surface
at certain temperature. Thermodynamical point of view, this means that only the region of the top of this fold with temperature lower the phase transition and of the bottom of this fold with temperature higher than the
phase transition is favoured.
Other region are thermodynamically unfavoured do not come into the stability discussion.
An interesting point is that on the top of the fold and just below the phase transition temperature we can
actually find a region where thermodynamically favourable but electrically unstable.
This region should play a central role in the canonical ensemble.
In \cite{Mateos:2007vc}, the similar problem was addressed.
The resolution of the problem in the D3/D7 system was that minimizing the free energy in the
grand canonical ensemble always picks out either a stable black hole embedding or Minkowski embedding.
Hence the system never suffer from the instability.
The question is whether the same is true for the D4/D6 system.
After explaining Figure \ref{BhMkPhaseTrans}, we come back to this question.

\FIGURE{
  \includegraphics[width=0.8 \textwidth]{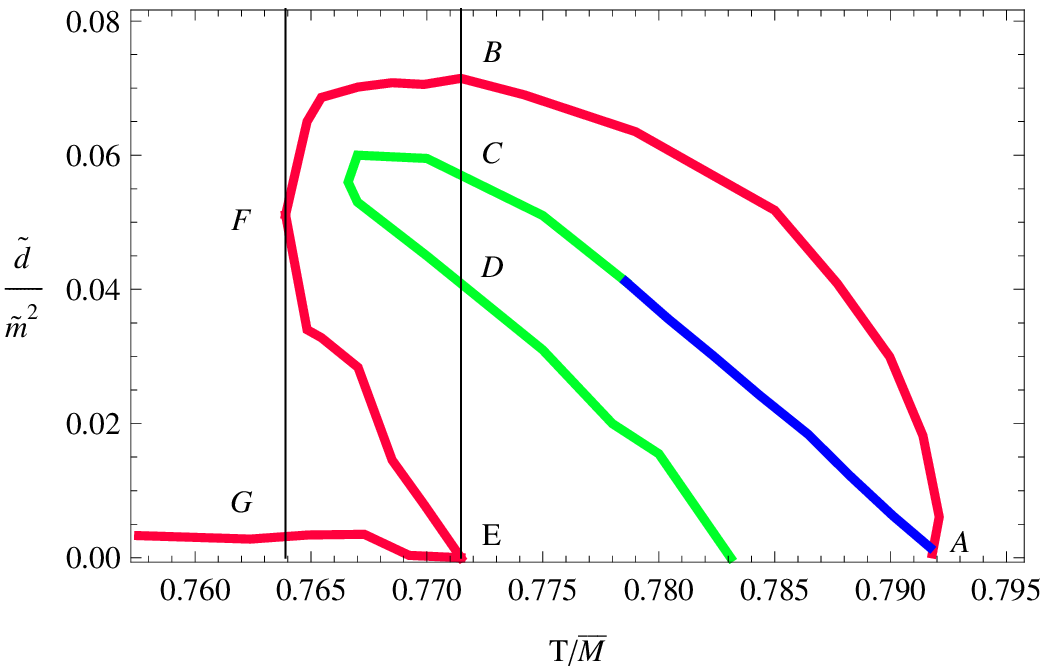}
\caption{Baryon density at the phase transition in the grand canonical ensemble (red)
The blue line of phase transitions identified in the canonical ensemble at fixed $n_b$.
The region enclosed by the green and blue curves corresponds to the unstable region in which
$(\partial \mu_q/\partial n_q)_{T}<0$.
} 
\label{BhMkPhaseTrans}
}

Figure \ref{BhMkPhaseTrans} shows the black hole to Minkowski phase transition, $\partial\mu_{q}/\partial n_b=0$, and the canonical phase transition line.
The blue line is the canonical phase transition line.
For fixed value of $d$, there is a black hole to black hole
phase transition at this line.
The green line is  a part of the ${\partial \mu_{q}\over \partial n_b}\bigr|_{T}=0$ line.
The region surrounded by green and blue line is the 
thermodynamically favourable but electromagnetically 
unstable region.
The red line is the mapping of the phase transition line in Figure \ref{muT} to $\tilde{d}/\tilde{m}^2-T/\bar{M}$ plane.
Let us explain this line more detail.
In Figure \ref{WvsTmu014}(b),  there is one phase transition 
from the black hole embedding to the Minkowski embedding at 
$T=0.7742$ and $\tilde{d}/\tilde{m}^2=0.069$.
This corresponds to the red line between A and B in Figure
\ref{BhMkPhaseTrans}. In this region there is only one value of 
$\tilde{d}/\tilde{m}^2$ for fixed $T/\bar{M}$ and at this point
the phase transition occurs from a Minkowski embedding to 
a black hole embedding, i.e., from a point at $\tilde{d}/\tilde{m}^2$=0 to a point on the red line with the same temperature. 

In Figure \ref{WvsTmu01650175}(a),
there are two phase transitions.
One is from the Minkowski embedding to the black hole
embedding at $T/\bar{M}=0.7649$ and 
$\tilde{d}/\tilde{m}^2=0.0034$.
The other one is from the black hole embedding 
at $T/\bar{M}=0.76703$ and $\tilde{d}/\tilde{m}^2=0.0285$
to the other black hole embedding
at $T/\bar{M}=0.76703$ and $\tilde{d}/\tilde{m}^2=0.07014$
This configuration corresponds to a region
between the two vertical black lines in Figure \ref{BhMkPhaseTrans}.
The first phase transition is from a Minkowski embedding
$\tilde{d}/\tilde{m}^2$=0 to a point on the red line between E and
G. Then the second phase transition occurs at a different temperature from a point on the red line between E and F to
a point on the red line between F and B.

In Figure \ref{WvsTmu01650175}(b), there is only one phase transition from the Minkowski embedding to the black hole embedding
at $T/\bar{M}=0.76$ and $\tilde{d}/\tilde{m}^2$=0.003.
This corresponds to the left part of G on the red line.

As in the D3/D7 system\cite{Mateos:2007vc}, the whole unstable region is surrounded by the red line.
Hence the grand canonical ensemble picks out either stable 
black hole embedding or Minkowski embedding; the thermodynamically favourable phases
discussed above are all stable. 
The unstable region appeared in the canonical ensemble is misidentified the ground state since 
we restricted our analysis to the homogeneous configurations.
The true ground state is an inhomogeneous phase of stable phases
\cite{Mateos:2007vc}.

The true ground state of the region surrounded by the red curve A-B and the $\tilde{d}/\tilde{m}^2=0$ axis is an inhomogeneous phase of stable black hole (on the red curve A-B) 
embedding and Minkowski embedding (on the $\tilde{d}/\tilde{m}^2=0$ axis).
The true ground state of the region surrounded by B-F-E-B is an inhomogeneous phase of stable black hole embedding
(on the red curve B-F) 
and another stable black hole embedding (on the red curve F-E).
The true ground state of the region surrounded by the red curve E-G and the $\tilde{d}/\tilde{m}^2=0$ axis, is an inhomogeneous
phase of stable black hole embedding (on the red curve E-F) and a Minkowski phase (on the $\tilde{d}/\tilde{m}^2=0$ axis).
Hence both in the grand canonical ensemble and the canonical ensemble, the unstable region
is thermodynamically unfavourable.

\section{Conclusion and Discussion}

We have seen that there are several common physical properties in the Dp/Dq systems
with finite charge density and chemical potential.
In the grand canonical ensemble there is a line of the first order phase transitions
from Minkowski to black hole embeddings found in \cite{korea,japan,karch3,Mateos:2007vc}.
The black hole embeddings cover high $T/\bar{M}$ or high $\mu_q/M_q$ region while
the Minkowski embeddings cover the whole $\mu_q/M_q$ region with $T/\bar{M}$ smaller
than $\sim T_{fun}$.
In the supersymmetric limit $T/\bar{M}\rightarrow 0$, $\mu_q/M_q$ always goes to one
for any value of $n_b$, i.e., the energy which is necessary to add one quark in the system
is equal to its bare mass. 
In this limit black hole embeddings exactly look like Minkowski embeddings.
There is a very thin and long spike stretching from Dq branes down to Dp branes.
In the dual field theory point of view\cite{Mateos:2007vn,spectre,hoyos}, the Minkowski embeddings correspond to the stable meson phase and
he spectral function consists of a series of delta-function-like peaks, i.e., resonances centred 
on mass eigenvalues. 
On the other hand, the black hole embeddings correspond to the unstable meson phase and 
the spectral function is continuous.
In the canonical ensemble, there is a line of first order phase transitions from 
black hole to black hole embeddings for charge density less than $n^* _b$ and 
above this critical density there is no phase transition.
Just below the phase transition temperature, there is an electrodynamically unstable region.
However this unstable region is not the true ground state and should be replaced by
an inhomogeneous phase. 

We also have seen that there is a difference between the D3/D7 and D4/D6 system.
In the D3/D7 system for fixed $\mu_q<M_q$, there is only one phase transition, from a Minkowski to
a black hole embedding.
On the other hand, in the D4/D6 system for certain range of $\mu_q<M_q$, there are two phase transitions, one is from a Minkowski to a black hole embedding
and the other one is from a black hole to another black hole embedding.
Following the investigation of \cite{spectre,Mateos:2007yp,erdmenger}, we can give a physical interpretation
of this black hole to black hole embedding phase transition in the dual gauge theory.
The feature of the spectral function is characterized by the poles of the corresponding retarded correlators
in the complex frequency plane.
In the Minkowski embeddings, the poles are on the real axis and the spectral function is, as mentioned above,
a series of delta-function-like peaks.
In the black hole embeddings, the poles are located apart from the real axis.
At lower temperature, the imaginary part of the poles are close to the real axis and the spectral function 
exhibits distinct peaks.
As temperature increases, the poles move away from the real axis and the spectral function becomes featureless.
Hence our black hole to black hole phase transition from lower temperature to higher temperature would
correspond to poles jumping from closer locations to the real axis to farther locations.
The peaks of the spectral function would become lower and the life time of the quasiparticles would become shorter. 
It would be very interesting to investigate this point concretely.
Lattice QCD studies\cite{lattice} suggest that meson bound states survive the deconfinement phase transition
and the bound states dissolve at $T_{fun}$.
As explained, Minkowski to black hole phase transition represents this meson dissolving phase transition.
A very interesting question is then what the phase transition in QCD corresponding to the
black hole to black hole phase transition is.
It would be an exotic phase transition if it exists in QCD.

Before ending the section, we comment one more common property in the D3/D7 and the D4/D6 systems.
As mentioned above, in the $T/\bar{M}\rightarrow 0$ limit, the chemical potential $\mu_q$ goes to $M_q$.
This result is reasonable in the case of free fermions with zero density.
However since our system is strongly interacting system with bosons and fermions,
the reason is not very clear.
In fact, the chemical potential is very close to the constituent mass at the phase transition points
not only at $T/\bar{M}=0$ but also for a broad range of temperature except the canonical phase transition region.
Hence the quarks behave as if they were `free' particles.
We have put `free' om quotes because the constituent mass takes into account all the corrections from 
the quarks and the adjoint plasma.
The constituent quark mass is defined by the energy of a fundamental 
string stretching from a probe D-brane to the horizon in a Minkowski embedding.
This constituent quark mass at $T/\bar{M}=0$ on Minkowski embeddings with
$n_q=0$ was calculated in \cite{Mateos:2007vn, Herzog:2006gh, Mateos:2007vc}
The induced metric of the fundamental string stretching from
a probe D-brane to the horizon in $R$ direction is
\beq
ds^2={1\over2}\left( {u_0R\over L}\right)^{3/2}\left[ -{f^2\over \tilf}dt^2\right]
+\left({L\over u_0R}\right)^{3/2}{\tilf^{1/3}\over 2^{1/3}}u^2 _0dR^2.
\eeq
The Nambu-Goto action for the fundamental string is
\bea
I&=&-{1\over 2\pi l^2 _{s}}\int ^{R_0} _{1}dRdt\sqrt{-\det P[G]}\\
&=&-{1\over 2\pi l^2 _{s}}\int dt{u_0\over 2^{2/3}}\left(\left(1+{1\over R^3 _0} \right)^{2/3}
R_0-2^{2/3} \right)
\eea
Identifying the constituent mass to minus the action per unit time
of this static configuration, we have
\beq
M_{c}={1\over 2\pi l^2 _{s}}{u_0\over 2^{2/3}}\left(\left(1+{1\over R^3 _0} \right)^{2/3}
R_0-2^{2/3} \right).
\eeq


%
\FIGURE{
\begin{tabular}{cc}
  \includegraphics[width=0.48 \textwidth]{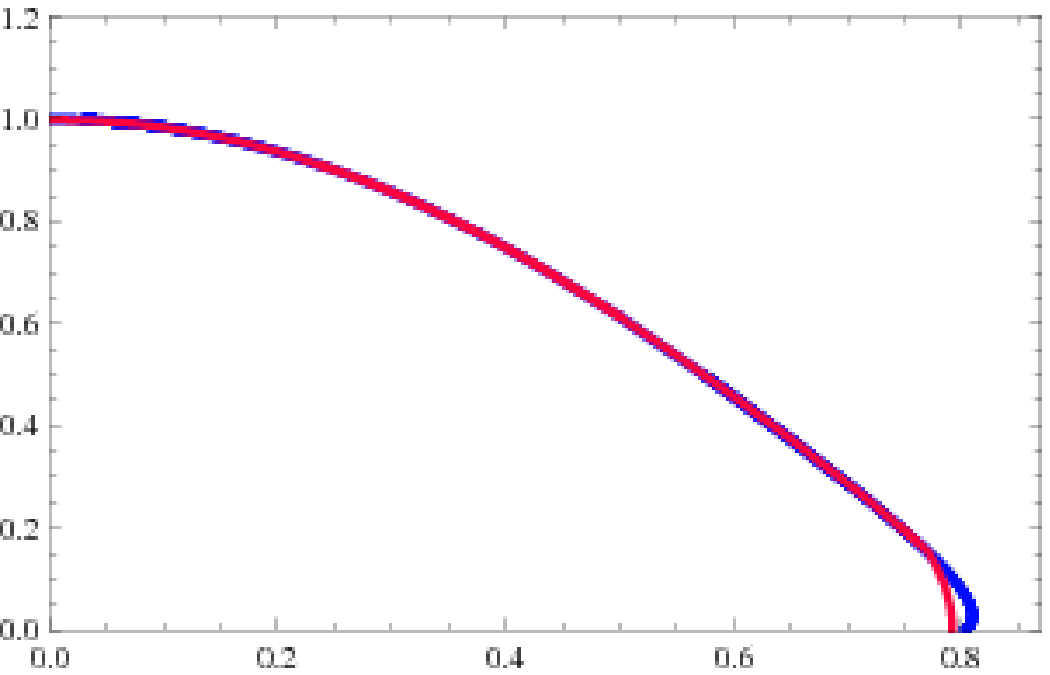}&
  \includegraphics[width=0.5 \textwidth]{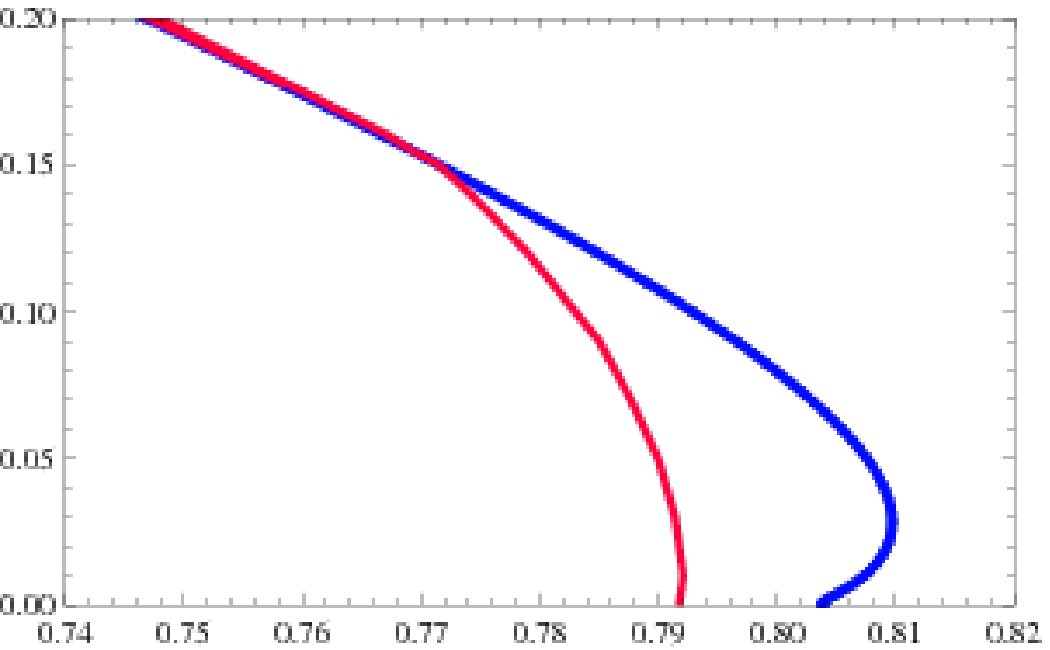}\\
(a)&(b) 
\end{tabular}
\caption{Comparison of the ratio $M_c/M_q$ (blue) with the ratio $\mu_q/M_q$ (red)
at which the phase transition from a Minkowski to a black hole embedding takes place.
The two curves essentially coincide on the scale of the Figure (a).} \label{constituent}
}

We can see that the constituent mass is almost identical to 
the chemical potential except near the canonical phase transition region.
This result is the same as that of the D3/D7 system\cite{Mateos:2007vc} and the similar result is obtained in \cite{japan}.
As mentioned in \cite{Mateos:2007vc}, this result is surprising because the quark density is much 
larger than the size of an individual quark;
\beq
n_q={2^{3/2}\over 3^3}N_fN_c g_{eff}(M_q) \left({\tilde{d}\over\tilde{m}^2}\right) n_{crit},
\eeq
where $g^2 _{eff}(M_q)=\lambda M_q=g^2 _5N_c M_q$ and $n_{crit}=\bar{M}^3$.
Structure functions\cite{holycow} or a quark's disturbance of the adjoint fields
\cite{riverside} in the D3/D7 system are well studied.
The mass spectrum of mesons in the Dp/Dq system is discussed in \cite{Myers:2006qr}.
Motivated by their works, we assume that the individual quark size is $\sim m_{gap}$.
Combining it with the relation ${\bar{M}/ m_{gap}}\simeq 0.233$\cite{Mateos:2007vn}, we define
the critical density as $n_{crit}=\bar{M}^3$ at which quarks start to overlap each other.
For the effective coupling, supergravity approximation is valid only in the region\cite{Itzhaki:1998dd}
\beq
1<<g_{eff}<<N^{4/3} _c.
\eeq
Hence although $\left({\tilde{d}/\tilde{m}^2}\right)$ may be of order $10^{-3}$,
$n_q$ is much larger than $n_{crit}$ and interactions would not be negligible.

%
%
%

\acknowledgments 
The author thanks Robert C. Myers, David Mateos and Rowan F.M. Thomson for useful
conversations.
This research was supported by Perimeter Institute for Theoretical Physics.  Research at Perimeter Institute is supported by the Government of Canada through Industry Canada and by the Province of Ontario through the Ministry of Research and Innovation.
The author also acknowledge support from a JSPS Research Fellowship for Young
Scientists.




\end{document}